\documentclass[runningheads]{svmult}

\usepackage{makeidx}   
\usepackage{graphicx}  
\usepackage{subeqnar}  
\usepackage{multicol}  
\usepackage{physprbb}  
\makeindex             

\begin{document}
\title*{Evidence Supporting the Universality of the IMF}
\toctitle{Evidence Supporting the Universality of the IMF}
%
%
\titlerunning{Evidence against IMF Variations}
%
\author{Gerry Gilmore}
\authorrunning{Gerry Gilmore}
%
%
\institute{Institute of Astronomy, \\
Madingley Road, Cambridge CB3 0HA, UK}

\maketitle              

\begin{abstract}
The stellar initial mass function (IMF) is an underlying distribution
function which determines many important observables, from the number of
ionizing photons in a population of some age and metallicity, through
the creation rate of various chemical elements, to the mass to light
ratio of a system. This significance, together with the empirical
difficulty to determine the IMF robustly, and the near complete lack of
any robust theoretical predictor, has allowed investigators freedom to
treat the IMF as a continuously variable parameterisation of
astrophysicists' ignorance of complexity. An ability to vary a parameter
in a model is not the same as a true variation in a physical system. A
more instructive approach is to use available data to constrain possible
variations, and thereby to allow identification of those other aspects
of an observed system whose understanding can be improved. Ideally, the
most sensitive physical variable, or its parameterisation, should be the
best constrained. A fundamental null hypothesis, which we defend here,
is that the IMF is a universal invariant function, so that all apparent
variations may be ascribed to other variables, and to irreducible
statistical sampling fluctuations.
\end{abstract}

\section{Introduction}

In the accompanying article, Eisenhauer \cite{E00gg} both introduces
this subject and provides a summary of many indirect evidences that
the IMF is seen to vary. Much of his evidence is based on the review
of John Scalo \cite{scalo98gg} in the recent conference ``The Stellar
Initial Mass Function'' \cite{gilmore98gg}. Scalo's review, while
compelling, is in disagreement with the remainder of the
contributions to that volume. Thus, it is worth noting immediately
that the concensus among expert opinion has changed dramatically in
recent years, as data have improved. Previously it was assumed that
the IMF was variable, and the task of observation was to quantify
those changes. One  hoped to identify some dominant parameter
(metallicity, environment,..), whose identification would in turn lead
star formation theory towards reality. The holy grail of explaining
(baryonic) dark matter in the universe was a further motivation in the early
days, though now accepted to be irrelevant to IMF studies: even in an
extreme interpretation of the data, (and lots of such analyses exist)
the stellar IMF is irrelevant to the dominant problem of dark matter.

The postulation of a universal IMF --- in its strongest form ---
states that the IMF is and has always been the same in all regions of
star formation everywhere; the frequencies of initial stellar masses
in any unbiased sample are always drawn from the same statistical
distribution function. 

A decade ago, the scientific null hypothesis was that the IMF should
vary, based on naive theory, and did vary, based on extant analyses of
observations. A second major open problem, identification of the
`missing mass', was certainly relevant to determination of the number of
low mass stars. Applying the long-standing scientific rule that any
explanation of one problem which incidentally explains a second is
progress, quantifying IMF variations became a reasonable default
experiment. As the data improved however -- the star count data has
extended faintward by some 20 magnitudes in the red/infrared during the last two
decades -- and as an appreciation of the physical complexity of star forming
regions observed only in integrated light arose, and, most importantly, as
the number of high-quality very detailed HST studies of star clusters
increased, the evidence in favour of apparent IMF variations has
decreased. The unexpected current result is that any variation in the
IMF is startlingly small. The observational situation has changed so
much that the conservative assumption has now inverted.

That is, the conservative assumption is now that the IMF is invariant,
and that apparent variations are due to unrecognised complexities in
the physical situation being analysed, and the statistics of small samples, 
both convolved with recognised
difficulties in the conversion of observables -- ionisation state,
metallicity, age range, stellar colour, ... into fundamental
parameters, such as mass. This null hypothesis is clearly extreme: zero
intrinsic variation in the IMF seems ab initio implausible, given the
diversity of physical conditions in which stars form. However, the
observations point to a situation in which variation in the IMF is a
second-order effect. 

Here we consider two complementary conservative questions: first, is
there any irrefutable evidence that the IMF is variable, and secondly,
do we know what the IMF is? I conclude that there is indeed no robust
evidence for variation, while there remains considerable (systematic)
uncertainty in quantifying the underlying (invariant) IMF.

\section{The IMF as a Function of Time}

Was the IMF different in the early Universe than it is today? A variety
of indications suggest such changes, primarily indirectly from attempts
to model the chemical evolution of the IGM, to explain galaxy cluster
abundances, and to explain the lack of very metal-poor stars in the
Galactic halo. All of these suffer inevitable uncertainty from the many
simplyfying approximations involved in modelling gas flows, accretion
histories, outflows, and so on.

\subsection{Chemical Abundance Constraints}

Although absolute chemical abundances remain poorly understood, some
much less indirect constraints are available from the relative
abundances of the chemical elements. The relative abundances of a subset
of the elements are determined purely by the high mass stellar IMF, and
the rate of star formation. The astrophysics of this analysis is
described in many published papers and need not be detailed here. (An
example of one of the many relevant reviews is \cite{GWK89gg}). 
The essential feature is that the ratio of the
elements created purely in high mass stars to those elements created
primarily in lower mass stars is manifestly a measure of the relative
numbers of such stars, and so the slope of the IMF.

\begin{figure}
\resizebox{\hsize}{!}{\includegraphics{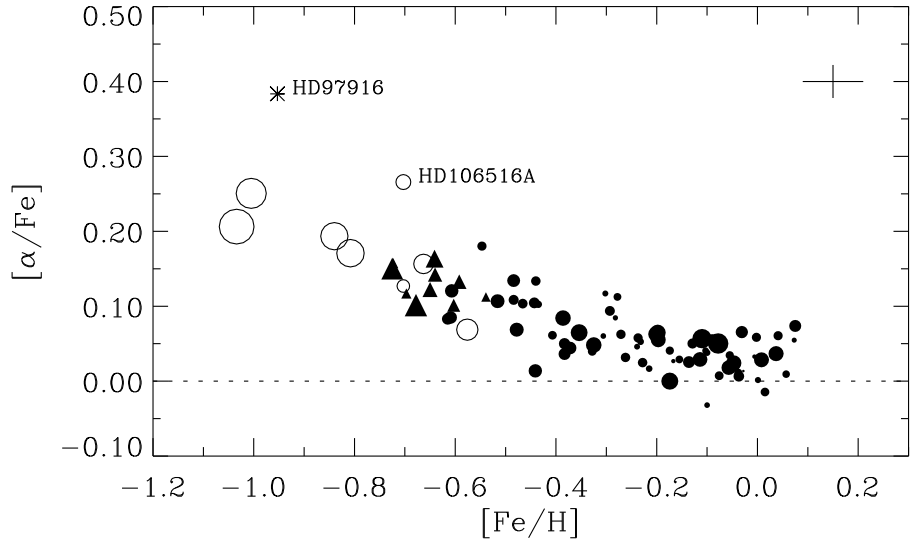}}
\caption{The mean $\alpha$ (Mg, Si, Ca and Ti) abundance as a
function of metallicity. The symbol
size is proportional to stellar age. (From  \cite{C00gg}). }
\end{figure}

The state of recent observational determinations of this ratio is
illustrated in Fig.~1, from Chen et al. \cite{C00gg}, who also
discuss the current uncertainties in understanding elemental production
sites. For present purposes however, the essential result is that
observational determinations, by many different authors, agree that
metal-poor stars show a systematic overabundance of the alpha elements,
created in high-mass stars, relative to the abundance of the iron-peak
elements, created in lower mass stars. The absolute value of the
overabundance thus provides a limit on the high-mass IMF slope at very
early times. Extant modelling (eg \cite{WG92gg})
limits any IMF variation relative to the present-day IMF to be less than
about 0.3 in slope, a maximum systematic change comparable to the
uncertainty in direct star-count determinations of the slope today.

Thus, modelling chemical element data provides no evidence for any
systematic change in the IMF slope over the whole age of the Universe.

\section{The IMF as a Function of Star Formation Rate}

It is very easy to expect the IMF to vary strongly in starburst regions.
The relevant observational suggestions are well reviewed by
Eisenhauer \cite{E00gg}. 
As so often is the case, one may however provide theoretical expectation 
to favour all possible outcomes. In particular, the concept of a Jeans
mass, and calculations of small perturbations from stability in a
self-gravitating system, are least likely to be relevant in a starburst,
where one must expect dramatic inhomogeneity and significant shock
energy. Analysis of observational data on current starbursts is
necessarily dependent on simplifying assumptions of the stellar age and
metallicity distributions, and is inevitably very sensitive to
the relative spatial distributions of stars, dust and the ISM,
together with any non-thermal energy sources which may be present.

There is one simple observational test, which does not probe the most
extreme high star formation rates, does probe most of the available
dynamic range, but is limited to low masses. That is to compare the
present-day IMF in the remnants of high and low star formation rate
systems. The remnant high-rate systems are globular clusters. These are
limited in duration of formation by the lack of a spread in abundances,
and additionally are seen today to be forming rapidly, and to be forming
in starburst galaxies. Globular clusters seem to form $10^6$ stars in
1Myr, for a rate of $1-10M_{\odot}yr^{-1}$. The remnant low star
formation rate systems are the dSph galaxies. The star formation rate
cannot have been high in these fragile galaxies. Quantitative
determinations of the rates, eg by Hernandez, Gilmore, and Valls-Gabaud
\cite{HGV00gg}, derive star formation rates of order $10^{-3}$ that
in galaxy disks today, and so some 3-7 orders of magnitude below those
of globular clusters. The difference in present stellar volume density
is similar.

\begin{figure}[ht]
\resizebox{\hsize}{!}{\includegraphics{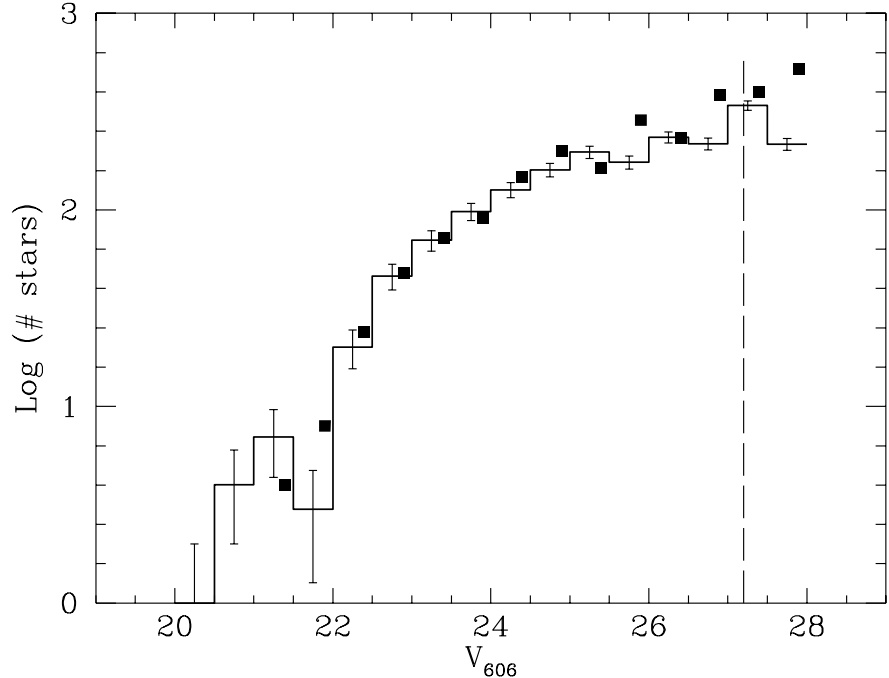}}
\caption
{Comparison between the UMi completeness-corrected luminosity
functions derived from the CMD with that of the globular cluster M92
(filled squares). The vertical
dashed lines indicate the 50\% completeness limits. The corresponding
mass range is $0.8 < M/M_{\odot} < 0.35$.
(From \cite{FWG99gg}).}
\end{figure}

A direct comparison of the stellar luminosity function in a globular
cluster and dSph galaxy has recently been obtained, by Feltzing, Wyse \&
Gilmore \cite{FWG99gg}. They compared the luminosity functions of
unevolved, low mass stars in the UMi dSph galaxy and in the globular
cluster M92. Both these systems are of similar age and metallicity, so
the luminosity function comparison is an IMF comparison. The systems
have star formation rates and present stellar density differing by about
6 orders of magnitude, and very different mass:light ratios. 
The results, shown in Fig.~~2, show very clearly
that there is no detectable IMF difference between these systems for
stars with masses in the range $0.8 < M/M_{\odot} < 0.35$.

\section{IMF Variations Between Similar Systems?}

Differential comparison between comparable systems with large numbers
of stars is the most sensitive way to identify real IMF variations. The
best case available, and the most studied, are the Galactic and Local
Group globular clusters. These contain large numbers of stars,
minimising statistical sampling errors (see below), and have little or
no internal metallicity range, minimising the need for reliable stellar
models (see below).

The mass-luminosity relation does of course change systematically with
me\-tal\-li\-ci\-ty, generating a systematic change in luminosity function with
metallicity. This is seen \cite{vonH96gg}, and is
to first order consistent with a constant IMF over the metallicity range
from $-2$dex to the Solar value. At a specific  metallicity one may in
principle provide more exact comparisons.

\begin{figure}
\resizebox{\hsize}{!}{\includegraphics{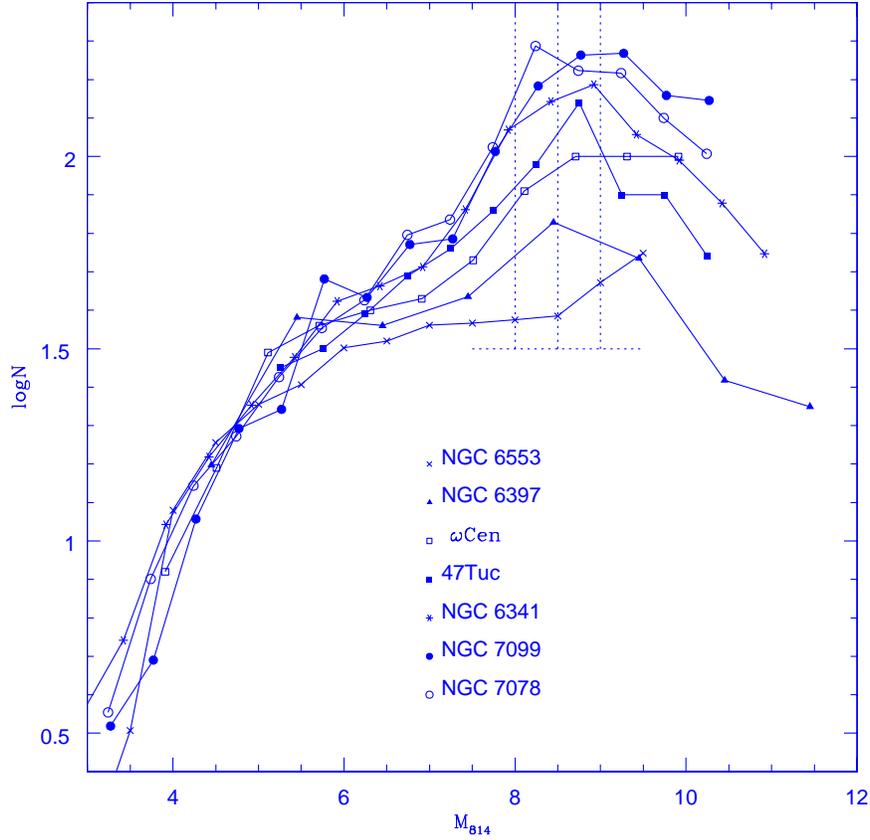}}
\caption{A collection of luminosity functions derived from HST
observations of Galactic globular clusters (Beaulieu etal in prep). 
Systematic differences
between clusters, even at the same metallicity, are evident. These
differences are all entirely consistent with a combination of the
metallicity dependence of the mass-luminosity relation, and with
different dynamical evolution between clusters. In particular, the
effects of the Galactic tidal field on internal cluster evolution can be
as large as the differences apparent here.}
\end{figure}

The extant HST cluster luminosity function data are
collected in Fig.~3, where large variations are apparent. Are these
real IMF variations? Systematic changes with time away from
 the primordial IMF of a
globular cluster are an inevitable consequence of  internal
dynamical evolution. Mass segregation must happen, as will stellar
merging and mass-dependent mass and star loss from
the cluster.  To date, modelling this rich mix of physical effects has
not been computationally feasible, so that highly simplified diffusion
modelling (Fokker-Planck) has been necessary. Such modelling is
necessarily inexact in the densest regions, where 2-body interactions
are dominant. The best available calculations of this type are still
unable to reproduce observed distributions of cluster
properties, so considerable caution is required.

\begin{figure}
\begin{center}
\resizebox{0.6 \hsize}{!}{\includegraphics{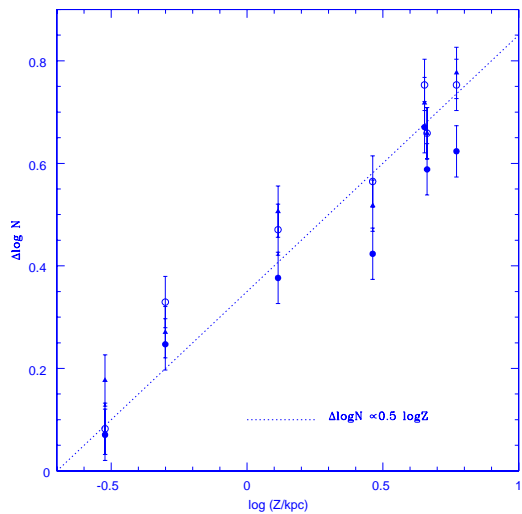}}
\end{center}
\caption{The differences in the globular cluster luminosity functions
shown in Fig.~3 ($\Delta$ log N stars at the LF maximum) 
plotted as a function of distance from the Galactic
Plane. This is a crude measure of the importance of tidal forces on
cluster evolution. The correlation suggest that tidal effects, which are
yet to be included completely in extant models, are as significant as
any real variation in the IMF between clusters.}
\end{figure}

Recent advances in N-body systems, especially GRAPE, are starting to
allow more realistic modelling, and hold great promise. In particular,
the important effects of cluster core dynamical evolution, and external
tides can be included reliably \cite{A99gg}. An indication
of the potential importance of these effects is seen in Fig.~4, which
plots the differences between cluster luminosity functions seen in
Fig.~3 as a function of distance from the Galactic Plane. This
distance is a crude measure of tidal forces, and suggests that these
additional factors are important. Preliminary modelling from extant
GRAPE facilities suggests that the range of luminosity function slopes
seen is (marginally) consistent with the effects of dynamical evolutions
which are feasible, assuming a universal IMF.

Until the next generation of such models are available, it is premature
to interpret the real differences in present day luminosity functions of
clusters as evidence for variation in the primordial IMF.

\section{Is the IMF Variable?}

Determination of the IMF from star count data is a complex challenge. In
addition to the obvious requirement for high-quality observational data
with reliable calibrations, there are many other factors in the
analysis. The most obvious among these include effects of stellar age,
multiplicity, and metallicity, mass segregation, sample incompleteness,
primordial and dynamical mass segregation.... These, plus all the many other
contributions to the Malmquist bias, were considered explicitely by
Kroupa, Tout \& Gilmore (KTG \cite{KTGgg}). Recently, Kroupa 
\cite{Kroupa2000gg}  has updated this
analysis, also including the effects of sampling noise. Sampling noise
is dominant in most published IMF determinations, which are studies of
young open clusters with few members. Kroupa's current ``best'' IMF is
shown in Fig.~5. Note that this IMF, which is very similar to that of
KTG, increases systematically to the hydrogen burning limit.

\begin{figure}
\begin{center}
\rotatebox{-90}{\resizebox{0.6 \textwidth}{!}
{\includegraphics{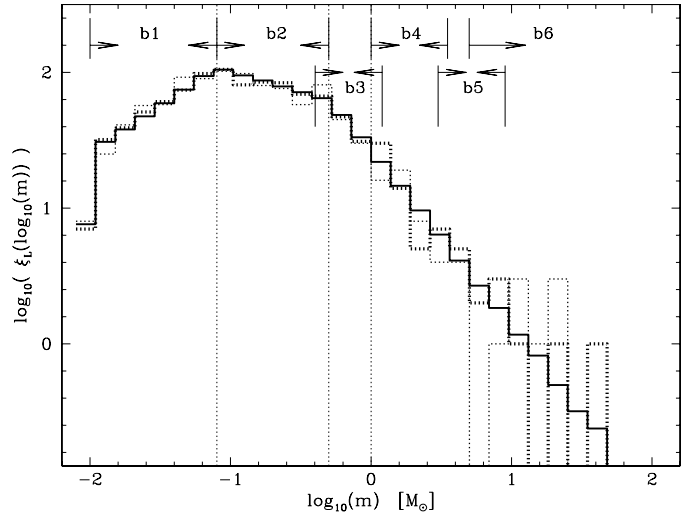}}}
\vskip 0mm
\caption
{\small{ The adopted logarithmic IMF for $10^6$ stars (solid
histogram). Two random renditions of this IMF with $10^3$ stars are
shown as the heavy and thin dotted histograms. The mass-ranges over
which power-law functions are fitted are indicated by the arrowed six
regions, while thin vertical dotted lines indicate
the masses at which the fitted slope  changes. (From \cite{Kroupa2000gg}). }}
\label{fig:mf_sv}
\end{center}
\end{figure}

To consider variations from this mean IMF in any observed sample, it is
easiest to consider $\alpha$, the gradient of this function, rather than the
function itself. This is illustrated in Fig.~6, the alpha-plot. 


%
\begin{figure}[ht]
\begin{center}
\includegraphics[width=0.6\textwidth,angle=-90]{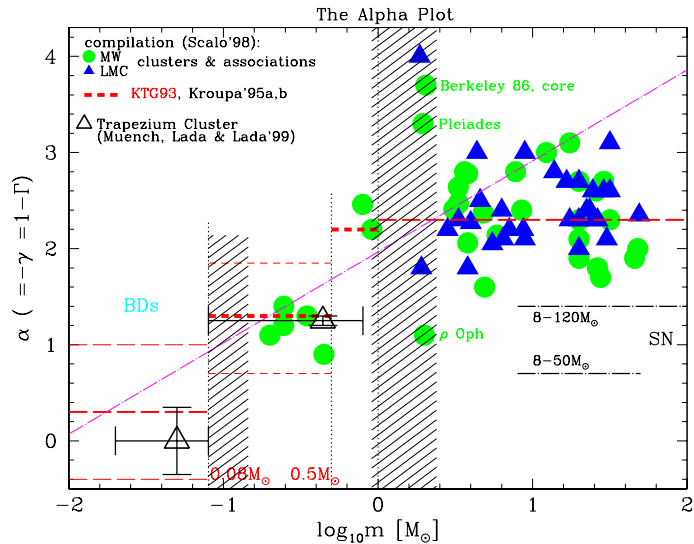}
\end{center}
\caption[]
{\small{ The {\it alpha-plot}.  The symbols show the compilation by Scalo
\cite{scalo98gg} of determinations of $\alpha$ over different mass ranges for
Milky-Way (MW) and Large-Magellanic-Cloud (LMC) clusters and OB
associations.  Unresolved multiple systems are not corrected for.  The
large open triangles (\cite{MLLgg}  from Orion Nebula
Cluster observations, binary corrections not applied) serve to
illustrate the present knowledge for $m<0.1\,M_\odot$.  The horizontal
long-dashed lines in the brown dwarf regime are the Galactic-field IMF
of Fig.~5 with associated approximate uncertainties. For
$0.08\le m \le 1.0\,M_\odot$ the thick short-dashed lines represent
the KTG single-star IMF (\cite{KTGgg}, which has
$\alpha_3=2.7$ for $m>1\,M_\odot$ from Scalo's \cite{scalo86gg}
determination. The long-dashed lines for $m>1\,M_\odot$ show the
approximate average $\alpha=2.3$, which is adopted in the
Galactic-field IMF.  The Miller \& Scalo \cite{ms79gg}
log-normal IMF for a constant star-formation rate and a Galactic disk
age of 12\,Gyr is plotted as the diagonal long-dash-dotted line.  The
long-dash-dotted horizontal lines labelled ``SN'' are those
$\alpha_3=0.70 (1.4)$ for which 50~\% of the stellar (including BD)
mass is in stars with $8 - 50 (8 - 120)\,M_\odot$.  The vertical
dotted lines delineate the four mass ranges highlighted in Fig.~5, and
the shaded areas highlight those stellar mass regions where the
derivation of the IMF is additionally complicated due to unknown ages,
especially for Galactic field stars: for $0.08<m< 0.15\,M_\odot$
long pre-MS contraction times make the conversion from an empirical LF to an
IMF dependent on the precise knowledge of the age, while for $0.8<
m<2.5\,M_\odot$ post-main sequence evolution makes derived masses
uncertain in the absence of precise age knowledge. Some published 
data are labelled by their star cluster names.
(From \cite{Kroupa2000gg}). }}
\label{fig:a_m}
\end{figure}


An important point is the extent to which the scatter of observational
points in the alpha-plot indicates real variation. In addition to the
systematic effects considered below, there is one important effect to
emphasise: statistical sampling noise, due to the finite and small
number of stars in a star-forming region. This has a disproportionate
effect at high masses, where the numbers are very small. The effect is
shown in Fig.~7,

\begin{figure}
\begin{center}
\rotatebox{-90}{\resizebox{0.6 \textwidth}{!}
{\includegraphics{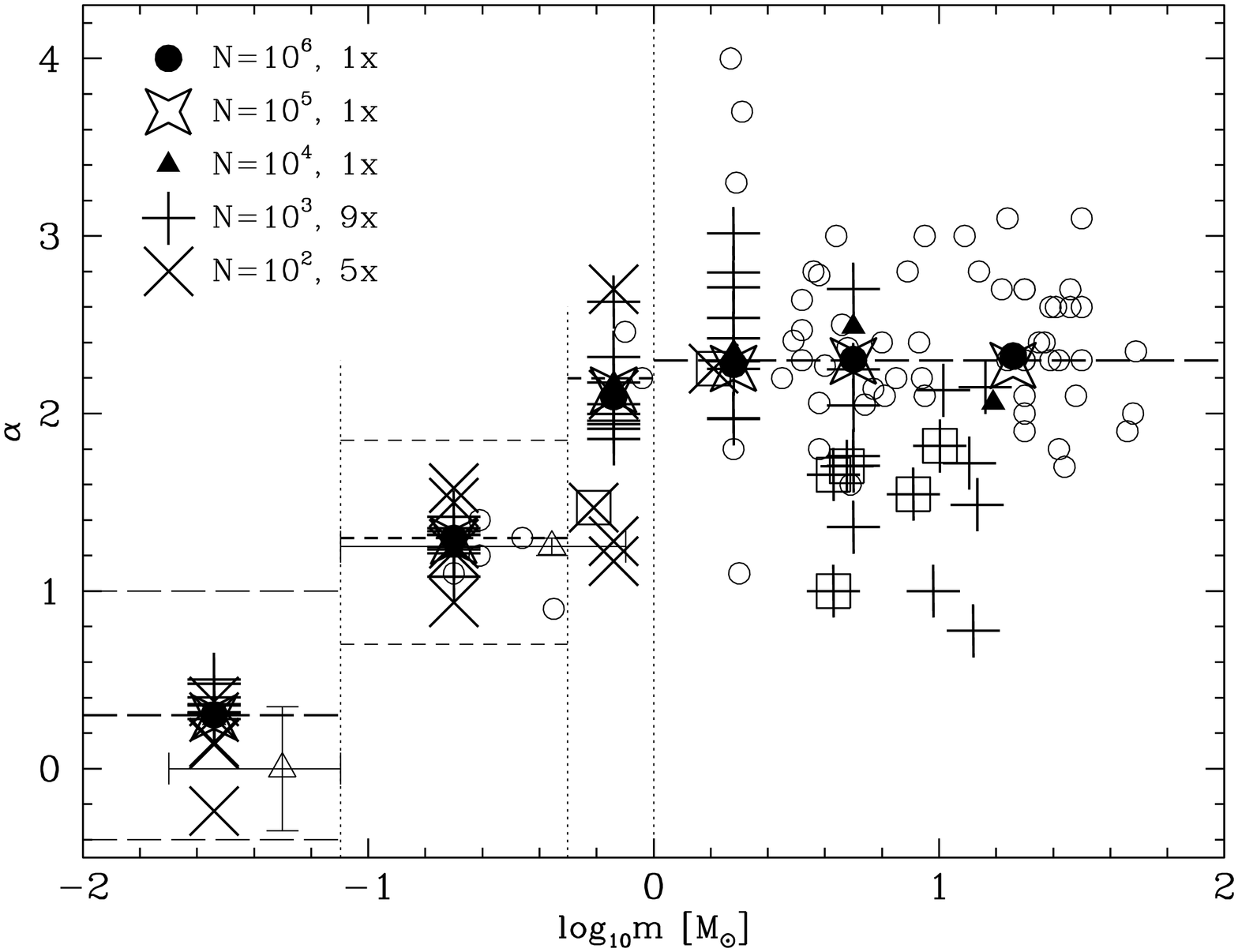}}}
\vskip 0mm
\caption
{\small{ Purely statistical sampling noise variation of the IMF slope
$\alpha$ in the six mass ranges of Fig.~5 for different observed star
numbers $N$ as indicated in the key. The open circles and open triangles
are as in Fig.~6. (From \cite{Kroupa2000gg}). }}
\label{fig:a_m_sv}
\end{center}
\end{figure}

Overall, this analysis illustrates that sampling and observational
effects, together with the many other contributions to Malmquist bias,
dominate available data: any direct evidence for IMF variation is not robust.

A further very important factor is illustrated in Fig.~8: even if one
has a large sample with accurate data, conversion of an observed
luminosity and derived effective temperature to source mass is not a
robust procedure. Extant stellar models are uncertain by an amount that
is large compared to deduced changes in the IMF \cite{CDPgg}.

\begin{figure}
\resizebox{\hsize}{!}{\includegraphics{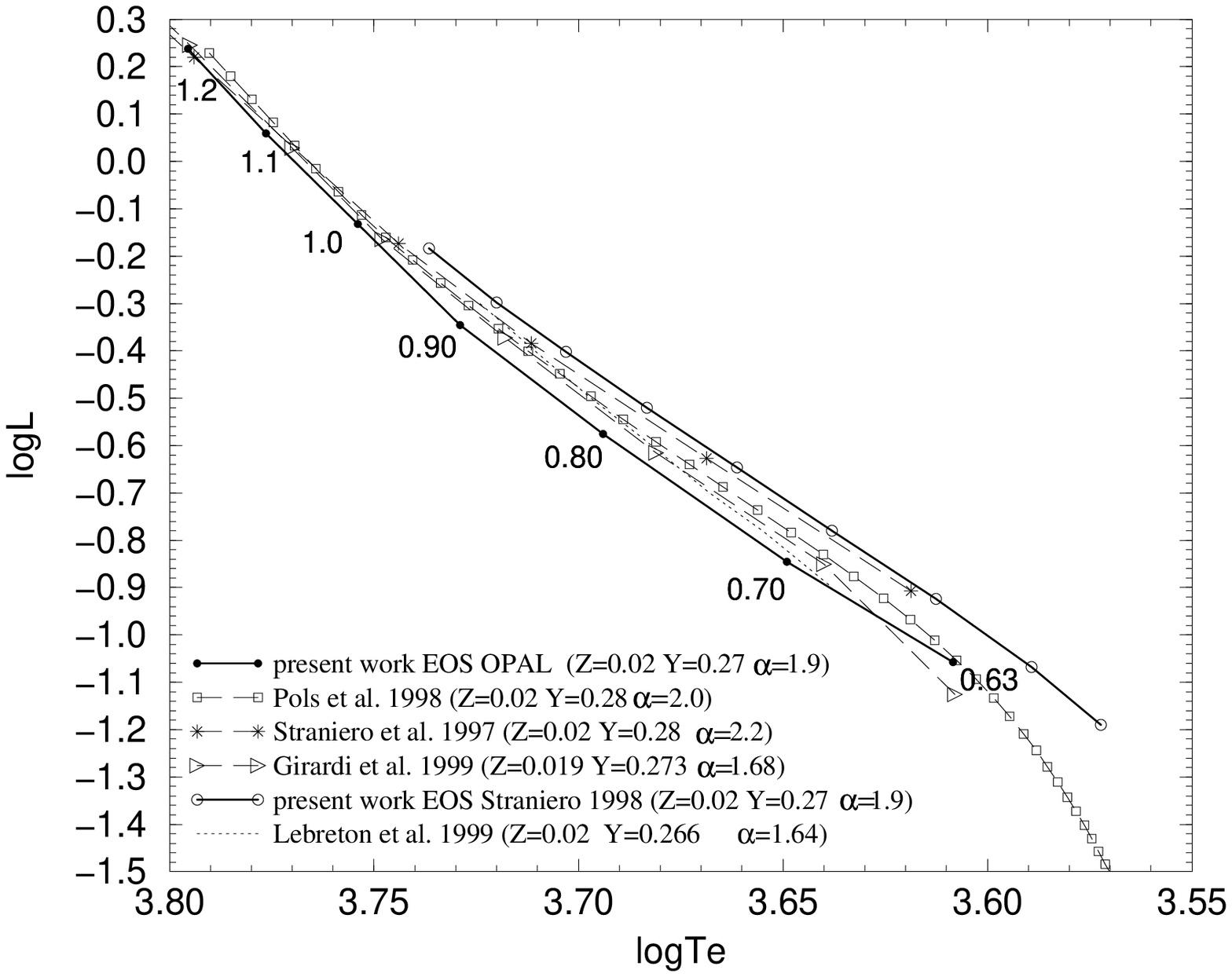}}
\caption{Location of the ZAMS in the HR diagram, for a variety of
recently published stellar models. The very considerable diversity of
model parameters which provide approximately the same slope in the HR
diagram illustrates the limits to which one can deduce fundamental
stellar parameters, such as mass, from an observation of luminosity and
temperature, given typical observational errors. (From  \cite{CDPgg}).}
\end{figure}

\section{Conclusions}

Variations of the stellar initial mass function have been reported for
all masses and in a large variety of stellar populations. We show that
in (almost) all cases the suggested variation is either dependent on
simplifying assumptions made in other important parameters, or is
dominated by sampling noise. Some direct evidence exists showing the IMF
has been invariant over 12\,Gyr, at all abundances from \mbox{-2dex} 
to at least
the Solar value, in systems with stellar densities covering 6 orders of
magnitude, and in systems with a range of star formation rates spanning
at least 6 orders of magnitude. This remarkable and unexpected
constancy, showing any real variation to be a second order, suggests the
physics underlying the IMF is dominated locally by the central limit
theorem, rather than one or a few dominant parameters.

\end{document}